\documentclass[twocolumn,aps,prc,superscriptaddress,floatfix,showpacs]{revtex4}

\usepackage{graphicx}
\usepackage{dcolumn}
\usepackage{bm}
\usepackage{color}
\begin{document}
\title{Sensitivity of neutron to proton ratio toward the high density behavior of symmetry energy in heavy-ion collisions}
\author{Sanjeev Kumar}
\author{Y. G. Ma\footnote{Author to whom all correspondence should be addressed:
ygma@sinap.ac.cn}} \affiliation{Shanghai Institute of Applied
Physics, Chinese Academy of Sciences, Shanghai 201800, China}
\author{G. Q. Zhang}
\author{C. L. Zhou}
\affiliation{Shanghai Institute of Applied Physics, Chinese
Academy of Sciences, Shanghai 201800, China} \affiliation{Graduate
School of the Chinese Academy of Sciences, Beijing 100080, China}
\date{\today}
\begin{abstract}
The symmetry energy at sub- and supra-saturation densities has
great importance for understanding the exact nature of asymmetric
nuclear matter as well as neutron stars, but it is poorly known,
especially at supra-saturation densities.  We will demonstrate
here whether or not the neutron-to-proton ratios from different kinds of
fragments can determine the supra-saturation behavior of
the symmetry energy. For this purpose, a series of Sn isotopes
were simulated at different incident energies using the isospin
quantum molecular dynamics (IQMD) model with either a soft or a
stiff symmetry energy. It is found that the
single neutron-to-proton ratio from free nucleons as well as
Light Charged particles (LCPs)
is sensitive to the symmetry energy, incident energy, and
isospin asymmetry of the system. However, with the double
neutron-to-proton ratio, this is true only for the free nucleons.
It is possible to study the high-density behavior of symmetry
energy by using the neutron-to-proton ratio from free nucleons.
\end{abstract}


\maketitle

\section{Introduction}
From the Bethe-Weizsacker mass formula, it is well understood
that the symmetry energy from bulk matter is the  difference
between the energy of pure neutron matter and pure symmetric
matter. Mathematically, it can we written as
\begin{equation}
E_{Sym}(\rho,\delta)~=~E(\rho,\delta=1)-E(\rho,\delta=0),
\end{equation}
where $\delta=\frac{\rho_n-\rho_p}{\rho_n+\rho_p}$ and
$\rho=\rho_n+\rho_p$. $\rho_n$, $\rho_p$, and $\rho$ are the neutron,
proton, and nuclear matter densities, respectively. The symmetry
energy has great importance in the dense matter existing in the
neutron stars, but only indirect information can be extracted from
astrophysical observations
\cite{Latt04}. It is also important in the quark gluon plasma
(QGP) and hadron gas (HG) phase \cite{Toro11}. The QGP and HG
phases existed in the early stage of the evolution of Universe
(about 15 billion years ago) and are inaccessible nowadays. It is difficult
to recreate these conditions, although numerous experiments
are occurring at the Relativistic Heavy Ion Collider (RHIC) and
the Large Hadron Collider (LHC)  \cite{Nagl11}. Heavy-ion
reactions,
during which matter goes through compression and expansion,
are considered to be the true testing ground for the hot and dense
matter phases. The nuclear equation of state (NEOS) and the density
dependence of the symmetry energy can be probed by some observables in
intermediate-energy heavy-ion collisions (HICs). The softness of
the NEOS has been well described in the literature in the last couple of
decades \cite{Aich91, Dani02}. However, the density dependence of
the symmetry energy, from the Coulomb
barrier to the deconfinement of nuclear matter, is a hot
topic in the present era. At sub-saturation
densities, the density dependence of the symmetry energy is studied by
observables such as the neutron-to-proton ratio, isotopic and isobaric
scaling, isospin diffusion, isospin fractionation and/or distillation,
and isospin migration. 
\cite{Li97,Zhan08,Sun10, Kuma11}. Recently, the MSU group
\cite{Zhan08} claimed the softness of the symmetry energy
at sub-saturation densities by using the double neutron-to-proton
ratio and isospin diffusion from two isotopic systems, $^{112}$Sn +
$^{112}$Sn and $^{124}$Sn + $^{124}$Sn at E = 50 MeV/nucleon.
In another study, again soft
symmetry energy 
was claimed by using the isospin diffusion for the same set of
reactions, but at E = 35 MeV/nucleon \cite{Sun10}. In a recent
study, soft symmetry energy is also favored for the same set
of reactions at E = 50 MeV/nucleon by using the neutron-to-proton
ratio \cite{Kuma11}. In all the studies, the problem of
sub-saturation density dependence of the symmetry energy seems to be
addressed to some extent; however, the uncertainties are still large
enough to justify the large amount of work that is under way in many
laboratories all over the world.

In contrast, the present status of supra-saturation density
dependence of the symmetry energy is quite uncertain and interesting.
The high-density behavior of the symmetry energy in the literature is
studied by using two important parameters: one is the yield ratio
parameter and second is the flow parameter. The yield ratio
parameter has been studied in term of single and double ratios of
neutrons to protons \cite{Li02,Li04,Li07}, single and double ratios
of $\pi^-/\pi^+$ \cite{Li02,Li03,Q05,Wolt09,Li09, Feng10,Zhang10},
the $\Sigma^-/\Sigma^+$ ratio \cite{Q05}, the $K^-/K^+$ ratio
\cite{Wolt09}, and isospin fractionation \cite{Li07}, while, the flow
parameter has been studied in terms of relative and differential
flows (single and double ratios) of neutrons to protons or $^3$H to
$^3$He  \cite{Li02,Yong06}, and in terms of the ratio \cite{Ruso11}
or difference
\cite{Cozm11} of neutron-to-proton elliptic flow. Before using the
$^3$H and $^3$He particle yield and flow ratios for the density
dependence of the symmetry energy at high incident energies, one must
check the production of these particles in the supra-saturation
density region, which is obtained during the highly compressed
stage only. However, the production of neutrons and protons occurs in
large amounts and can explain the high density dependence of
symmetry energy with great accuracy. favorable results with
neutron and proton elliptic flow at E = 400 MeV/nucleon were also
observed in 2011. In one of the studies, the softness of the symmetry
energy with $\gamma_i=0.9$ is predicted by comparing the FOPI
collaboration data
with the neutron-to-proton elliptic flow ratio \cite{Ruso11}. In
the same year, Cozma {\it et al.} \cite{Cozm11} predicted the
softness of symmetry energy with $x$ = 2 by comparing the FOPI
data with the neutron-to-proton elliptic flow difference. Even
then uncertainty lies in the results, in terms of determination of
symmetry energy: in the first study, symmetry energy is momentum
independent, while in later one it is from momentum-dependent
interactions. Moreover, the studies were limited to only 400
MeV/nucleon.

Let us examine some interesting features from the ratio parameters at
supra-saturation densities. All the ratio parameters show
sensitivity to the symmetry energy. In the literature, it is also
claimed that $K^0/K^+$ and $\Sigma^-/\Sigma^+$ have more
sensitivity than $\pi^-/\pi^+$ \cite{Q05,Wolt09}.
The sensitivities of all the parameters is checked in term of
transverse momentum and rapidity distribution dependence
\cite{Li02,Li04,Li07,Q05,Zhang10,Yong06}, while pion and kaon
ratio studies are extended with the isospin asymmetry of the system
and the incident energy \cite{Li03,Wolt09,Li09,Feng10}. In recent
years, when the pion ratio has been compared with the FOPI data by using the
two well known models IBUU04 and ImIQMD, in terms
of isospin asymmetry and incident energy, the predictions for the symmetry energy
are found to be totally opposite. ImIQMD predicts stiff
symmetry energy ($\gamma_i=2$) \cite{Feng10}, while IBUU04
predicts soft symmetry energy (x=1) \cite{Li09}.

In the present era, the $\pi^-/\pi^+$ ratio is supposed to be a strong
candidate for predicting the high-density behavior of symmetry energy.
Just as for  $\pi^-$ and $\pi^+$, neutrons and protons are
also produced in large amounts up to 1 GeV/nucleon. Even around 400
MeV/nucleon, the production of neutrons and protons is greater
than the production of pions. Unfortunately, the neutron-to-proton ratio
parameter in most studies is restricted only with the transverse
momentum and kinetic energy dependencies \cite{Li02,Li04,Li07}. To
draw a fruitful conclusion in the near future, first, it is
very important to study the isospin asymmetry and incident energy
dependencies of the neutron-to-proton ratio, just as in the recent
$\pi^-/\pi^+$ study, and then compare the sensitivity to the symmetry
energy from both ratios, as the pion ratio results were recently
compared with the FOPI experimental findings. Second, one  has
to avoid choosing randomly any type of fragment to study the
supra-saturation density dependence of symmetry energy. For this,
it is important to check whether or not a particular type of fragment
is formed in the region $\rho>\rho_0$, which is
simple when one addresses the sub-saturation density dependence of
symmetry energy. Finally, with increasing incident energy, the time
evolution of the density has a different trend at two extremes: one at
the time of maximum compression and the second at the freeze-out time
(t = 200 fm/c). It also becomes interesting to see the different
stiffnesses of the symmetry energy  dependence of the neutron-to-proton
ratio for incident energies at the time of maximum compression
and at freeze-out time.

In the concluding
remarks of this paper, we have tried to address the following
goals:
\begin{itemize}
 \item{To check the sensitivities of different kind of fragments
to the high-density behavior of symmetry energy.}
 \item{To check the behavior of the neutron-to-proton ratio at the time of maximum compression and at freeze-out time.}
\item{To check the isospin asymmetry and incident energy
dependences of single and double neutron-to-proton ratios from
different neutron-rich systems to the high-density behavior
of symmetry energy, and then compare the sensitivity of symmetry
energy from the neutron-to-proton ratio with that from the pion ratio. This study is
similar to recent studies using the pion ratio
\cite{Li03,Wolt09,Li09,Feng10}. }
\end{itemize}

For the present study, the isospin quantum molecular dynamics (IQMD)
model is used to generate the phase space of nucleons, which is
discussed in Sec. II. The results are discussed in sec. III,
and we summarize the results in sec. IV .
\section{Methodology: ISOSPIN QUANTUM MOLECULAR DYNAMICS MODEL}
\label{IQMD}

In the IQMD model \cite{Hart98, Kuma10}, nucleons are represented by wave packets,
just as in the QMD model of Aichelin \cite{Aich91}. These wave packets of the
target and projectile interact via the full Skyrme potential energy, which is
represented by $U$ and is given as:
\begin{equation}
U~=~U_{\rho} + U_{Coul}\cdot
\end{equation}
Here $U_{Coul}$ is the Coulomb energy and  $U_{\rho}$ originates
from the density dependence of the nucleon optical
potential, and is given as
\begin{equation}
U_{\rho}~=~\frac{\alpha}{2}\frac{\rho^2}{\rho_0}~+~\frac{\beta}{\gamma+1}
\frac{\rho^{\gamma+1}}{\rho_0^{\gamma}}+
E^{pot}_{Sym}(\rho)\rho\delta^2 .
\label{equation2}
\end{equation}

The first two of the three parameters of Eq.~\ref{equation2}
($\alpha$ and $\beta$) are determined by demanding that, at normal
nuclear matter densities, the binding energy should be equal to 16
MeV  and the total energy should have a minimum at $\rho_0$. The
third parameter $\gamma$ is usually treated as a free parameter.
Its value is given in term of the compressibility:
\begin{equation}
\kappa~=~
9\rho^{2}\frac{\partial^{2}}{\partial\rho^{2}}\left(\frac{E}{A}\right)~~.
\end{equation}
The different values of compressibility give rise to soft and hard
equations of state. The soft equation of state is employed in the
present study with the parameters $\alpha~=~-356$ MeV,
$\beta~=~303$ MeV, and $\gamma$ = 7/6, corresponding to an isoscaler
compressibility of $\kappa~=~200~$ MeV. In the  third term
$E^{pot}_{Sym}$  is the potential part of the symmetry energy,
which is adjusted on the basis of calculations from the
microscopic or phenomenological many-body theory, having the form
\begin{equation}
E^{pot}_{Sym}~=~\frac{C_{s,p}}{2}\left(\frac{\rho}{\rho_0}\right)^{\gamma_i}
.
\end{equation}
 Here $C_{s,p}~=~35.19~$MeV, parameterized on the basis of the experimental
value of the symmetry energy, is known as the symmetry potential energy coefficient.
On the basis of the $\gamma_i$ value, symmetry energy is divided into two types
with $\gamma_i~=~0.5$ and $\gamma_i~=~1.5$, corresponding to soft and stiff symmetry
energies, respectively.

The total symmetry energy per nucleon employed in the simulation
is the sum of the kinetic and potential terms and is given as
\begin{equation}
E_{Sym}(\rho)~=~\frac{C_{s,k}}{2}\left(\frac{\rho}{\rho_0}\right)^{2/3}~+~E^{pot}_{Sym},
\end{equation}
where $C_{s,k}~=~
\frac{\hbar^2}{3m}\left(\frac{3\pi^2\rho_0}{2}\right)^{2/3}~\approx~25$
MeV is known as the symmetry kinetic energy coefficient. The kinetic
symmetry energy originates from the Fermi-Dirac distribution
\cite{thesis}.

Finally, we get a density and isospin-single particle potential in nuclear matter as follows:
\begin{eqnarray}
V_{\tau}(\rho,\delta)&=& \alpha\left(\frac{\rho}{\rho_0}\right)+
\beta\left(\frac{\rho}{\rho_0}\right)^{\gamma}+
E^{pot}_{Sym}(\rho)\delta^2 \nonumber\\
& &+\frac{\partial E^{pot}_{Sym}(\rho)}{\partial\rho}\rho\delta^2+
E^{pot}_{Sym}(\rho)\rho\frac{\partial\delta^2}
{\partial\rho_{\tau,\tau^\prime}}.
\end{eqnarray}
Here $\tau \neq \tau^\prime$, $\frac{\partial\delta^2}{\partial\rho_n}~=~
\frac{4 \delta \rho_p}{\rho^2}$, and $\frac{\partial\delta^2}
{\partial\rho_p}~=~\frac{-4 \delta \rho_n}{\rho^2}$.
The potential also depends on the momentum-dependent interactions, which are optional in the IQMD model.

Note that the $\gamma$ used in the determination of the equation of
state and $\gamma_i$ used in the determination of symmetry energy
are different parameters. The interesting feature of symmetry
energy is that its value increases with decreasing
$\gamma_i$ at sub-saturation densities, while the opposite is
true at supra-saturation densities. In other words, soft symmetry
energy is more pronounced at sub-saturation densities, while
stiff symmetry energy is more pronounced at supra-saturation densities.

In the calculations, we use the isospin-dependent in-medium cross section in the collision
term and the Pauli blocking effects as in the QMD model \cite{Aich91}. The cluster yields
are calculated by means of the coalescence model, in which particles with relative
momentum smaller than $P_{Fermi}$ and relative distance smaller than $R_{0}$ are
coalesced into a one cluster. The value of $R_{0}$ and $P_{Fermi}$ for the present
work are 3.5 fm and 268 MeV/c, respectively.

\section{Results and Discussions}

The neutron-to-proton ratio is among the first observables that
was proposed as a  possible sensitive probe for symmetry energy
prediction at sub-saturation densities \cite{Li97, Zhan08};
however, some studies are also performed using the rapidity
distribution and transverse momentum dependencies at
supra-saturation densities. In this article, the sensitivities of
free nucleons, light charged particles (LCPs, having charge
number between 1 and 2), and intermediate mass fragments (IMFs,
having charge number between 3 and $Z_{Total}/6$) to
the high density behavior of symmetry energy are checked,
providing the results of incident energy and isospin asymmetry
dependencies of single and double neutron-to-proton ratios with the
high-density sensitive fragments.

To perform the  study, thousand of events are simulated for the
isotopes of Sn, namely $^{112}$Sn + $^{112}$Sn, $^{124}$Sn +
$^{124}$Sn and $^{132}$Sn + $^{132}$Sn between incident energies of 50
and 600 MeV/nucleon at semicentral geometry by using the soft and stiff symmetry energies
of $\gamma_i~=~0.5$ and 1.5, respectively.
 As discussed earlier, a soft equation of state with an isospin-
dependent nucleon-nucleon (NN) cross section of
$\sigma_{med}~=~\left(1-0.2\frac{\rho}{\rho_0}\right)\sigma_{free}$
is employed. The incident energy and isospin asymmetry dependences
of single and double neutron-to-proton ratios, just as for the
$\pi^-/\pi^+$ ratio \cite{Li09,Feng10}, are considered as a point
of importance in the present study. The single ratio is just the
ratio of neutrons to protons and is represented in the study by
$R(N/Z)$, while
 double ratio is the ratio of the single ratios of any two isotopes of Sn. In order
to study the systematics of the isospin effects, the single ratio
of the isotope with a greater number of neutrons is always mentioned in
the numerator when the double ratio is calculated. Mathematically,
the double ratio is represented by $DR(N/Z)$ and is given as
\begin{equation}
DR(N/Z)~=~\frac{R(N/Z)^{neutron~rich}}{R(N/Z)^{neutron~weak}} \cdot
\end{equation}
\begin{figure}
\vspace{-0.8cm}
\includegraphics[width=90mm]{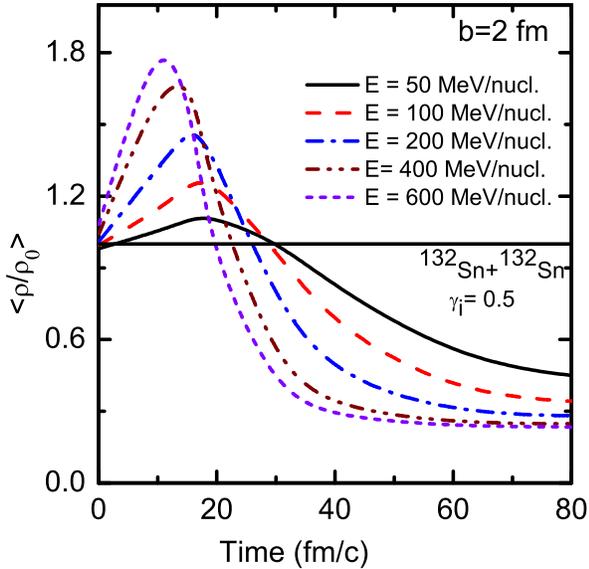}
\vspace{-4.5cm}\caption{\label{fig:1}(Color online) Time evolution of the average
density for the $^{132}$Sn + $^{132}$Sn reaction with the soft
symmetry energy ($\gamma_i = 0.5$) at semicentral geometry.
 The different lines represent different incident
energies ranging from 50 to 600 MeV/nucleon.}
\end{figure}

To predict the high-density behavior of symmetry energy, the very
first point is to understand the time evolution of the average density
at different incident energies. With increasing incident
energy, the density will be expected to be greater than the normal nuclear
matter density in the most compressed region. As we know, the
density of the environment surrounding the nucleons of a fragment
plays a crucial role in determining the physical process behind its
formation. In Fig. \ref{fig:1}, we display the average density
$\left<\rho/\rho_0\right>$ reached in a typical reaction as a
function of time at different incident energies for $^{132}$Sn +
$^{132}$Sn by using the soft symmetry energy $\gamma_i~=~0.5$. The
average nucleon density is calculated as \cite{Aich91}
\begin{eqnarray}
\langle \rho \rangle ~=~\langle\frac{1}{A_T+A_P}&
&\sum_{i=1}^{A_T+A_P}\sum_{j=1}^{A_T+A_P}
\frac{1}{(2\pi L)^{3/2}}\nonumber\\
& &\cdot exp[-(\vec{r_i}-\vec{r_j})^{2}/2L]\rangle,
\end{eqnarray}
with $\vec{r}_i$ and  $\vec{r}_j$ being the position coordinates
of the $i^{th}$ and $j^{th}$ nucleons, respectively.

 As we have expected,
with increasing of incident energy, the density is found to
increase in the compression zone.
 Interestingly, at lower beam energies, the maximum density
reached is lower and the reaction time is longer. With increasing
incident energy, the life-time of the high-density nuclear
matter gets shorter due to instability. For
example, at b = 2 fm the average density reaches a maximum and
is close to normal nuclear matter density at $t$= 18 and 33
fm/c, respectively, for E = 50 MeV/nucleon; but for the case of E
= 600 MeV/nucleon, the respective times are 10 and 20 fm/c.
This means that the difference between the two times
is almost 15 fm/c at E = 50 MeV/nucleon, while it is only 10 fm/c at
E = 600 MeV/nucleon. This clearly indicates that the matter shows
high-density behavior only for a small time interval, which decreases
with increasing incident energy. Since we are
interested in the sensitivities of different kinds of fragments and
their neutron-to-proton ratios, only those fragments that
lie in the high-density region ($\rho>\rho_0$) will be sensitive
to the high-density behavior of the symmetry energy.
\begin{figure}
\vspace{-0.8cm}
\includegraphics[width=90mm]{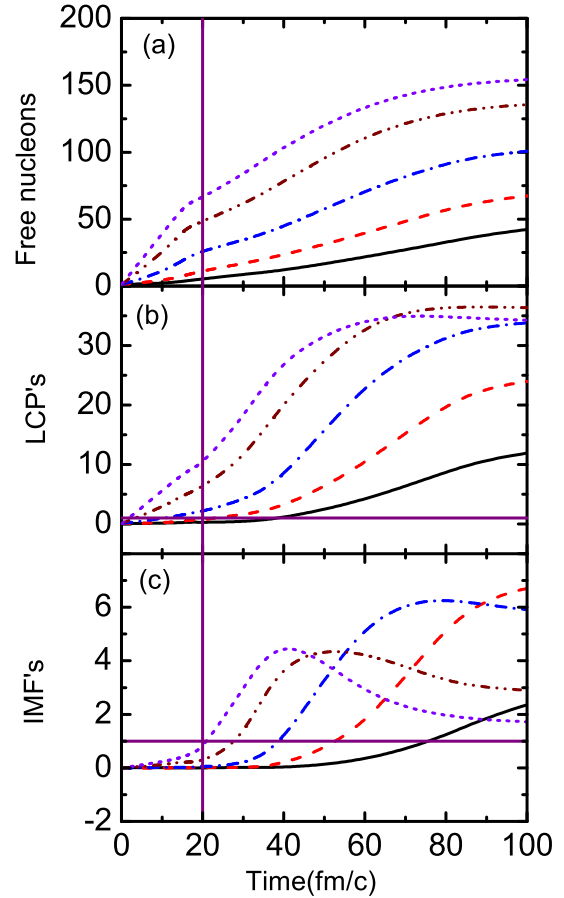}
\caption{\label{fig:2}(Color online) Time evolution of free
nucleons (top), LCPs (middle), and IMFs (bottom) at semicentral
geometry for $^{132}$Sn +
$^{132}$Sn using the soft symmetry energy ($\gamma_i=0.5$). The
different lines have the same meaning as in Fig.~\ref{fig:1}.
The vertical line in each panel represents our time limit before which
the system can be in the supra-saturation region (refer to Fig.
1).}
\end{figure}

To check the sensitivities of different kind of fragments, in Fig.
\ref{fig:2} we display time evolution of free nucleons (top),
LCPs (middle), and IMFs (bottom) at semicentral
 geometry  for incident energies
ranging from 50 to 600 MeV/nucleon. The behavior for all kinds of
fragments is consistent with the results in the literature
\cite{Kuma08}. The production of free nucleons increases with
incident energy, and LCP production  decreases after 400
MeV/nucleon. In Ref.~\cite{Kuma10}, LCP production is
correlated with the nuclear stopping and is also found to have a
maximum at 400 MeV/nucleon. IMF
production is found to decrease after 100 MeV/nucleon.
 This is due to the different origin of the production
of IMFs as compared to free and LCPs. For more details about
the incident energy  dependence of IMFs, see
Ref.~\cite{Kuma08}.

Our main task is to check the sensitivities of the fragments in the high-density
region. For this, we apply the limit that at least one
particle must be produced before the time 20 fm/c, because,  in an
average, after that time the density becomes lower than normal nuclear
matter density for all the incident energies under consideration.
The free nucleons are highly sensitive at all the energies.
 This is not true for LCPs and IMFs. LCPs are produced in this region
only after the
incident energy reaches 200  MeV/nucleon. In contrast, no IMFs
are produced in the supra-saturation density region.
This means that IMFs are not so sensitive to the
high-density dependence of symmetry energy; however, they can be
used at sub-saturation and saturation densities \cite{Zhan08}. Interestingly, the
single neutron-to-proton ratio from IMFs is found to change with
the incident energy (not shown here), but this is mainly due to
Coulomb interactions. Here we conclude that the neutron-to-proton
ratio from free nucleons as well as LCPs can act as a
probe of the high-density behavior of the symmetry energy.

One more interesting observation is obtained from Fig.
\ref{fig:1}. With increasing incident energy,
the time evolution of the density is exactly opposite during the
compression and expansion stages.
That is, in the expansion stage the average density is
found to decrease with increasing incident energy, which
was earlier increasing in the compressed zone. Now, we have two
aspects of the basis of the time evolution of density: one is the
compressed-zone time and second is the freeze-out time.
Interestingly, if the density behavior is opposite at the two
times, then it would supposedly affect the magnitude of the symmetry
energy as well as its effect on the nuclear matter during the
whole  time evolution.
\begin{figure}
\vspace{-1.8cm}
\includegraphics[width=90mm]{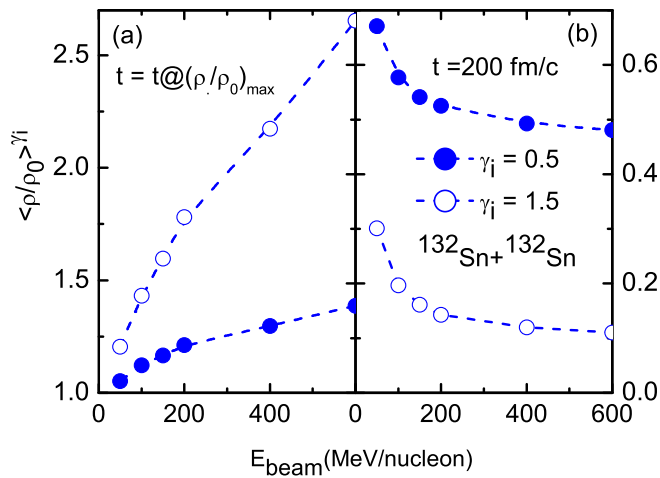}
\vspace{-4.8cm}\caption{\label{fig:3}(Color online) Excitation function of
$(\rho/\rho_0)^{\gamma_i}$, which is proportional to $E_{Sym}$, at
semicentral geometry. The left panel is at the time of maximum
compression, while the right one is at the freeze-out time. Solid and
open circles with a dashed line represent the contributions of the
soft ($\gamma_i$=0.5) and stiff ($\gamma_i$=1.5) symmetry
energies, respectively,  for the system $^{132}$Sn + $^{132}$Sn.}
\end{figure}

To see the virtual change in the symmetry energy due to the change
in the density, we display in Fig. \ref{fig:3} the incident energy dependence of
$(\rho/\rho_0)^{\gamma_i}$, which is
proportional to the symmetry energy, at the time of the maximum
compression (left panel) and at the freeze-out time (right panel).
At the time of maximum
compression, the symmetry energy rises with the increasing
incident energy (increase in density). As the density is more than
the normal nuclear matter density in this region, the stiff symmetry
energy is  stronger than the soft one. With increasing
incident energy (increase in density), the stiff symmetry energy is
changing drastically, while, the soft symmetry energy shows little
change. This exactly coincided with the ideal picture of density
dependence of the symmetry energy.  On the other hand, if we look
at the energy dependence of $(\rho/\rho_0)^{\gamma_i}$ at $t$
= 200 fm/c, the situation is totally different. The symmetry
energy is found to decrease with increasing incident energy
(decrease in density). Now the density is lower than the
normal nuclear matter density, so the magnitude of the soft symmetry
energy is greater than that of the stiff symmetry energy.
In other words, the
supra-saturation (sub-saturation) density region is more neutron rich with
$\gamma_i=1.5$ ($\gamma_i=0.5$). The effect from the sub- and supra-saturation
density behaviors of symmetry energy will compete and
contribute in the final observables. Due to the different behavior
of density at different times, it is important to
observe the isospin effects at the time of maximum compression and
at the freeze-out time to understand the high-density behavior of
symmetry energy. For this purpose, in the coming sections,
the incident energy and
isospin asymmetry dependencies of the single and double neutron-to-proton
ratios from free nucleons and LCPs are analyzed.
\begin{figure}
\vspace{-1.0cm}
\includegraphics[width=90mm]{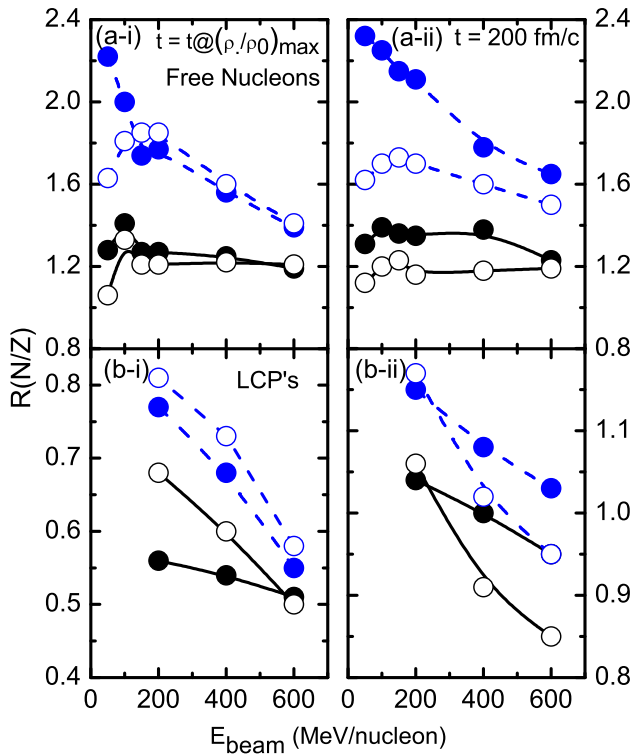}
\vspace{-1.5cm}\caption{\label{fig:4}(Color online) Excitation function of the single
neutron-to-proton ratio for free nucleons (top panel)
and LCPs (bottom panel) at semicentral geometry. The left and right panels are
the same expect they are at the time of maximum compression and at the freeze-out
time, respectively. Solid and open circles represent the soft and
stiff symmetry energies, respectively. Solid and dashed lines
corresponds to the systems of $^{112}$Sn + $^{112}$Sn and
$^{124}$Sn + $^{124}$Sn, respectively. }
\end{figure}

\subsection{Incident energy and isospin asymmetry dependencies of single neutron to proton ratio}
\begin{figure}
\vspace{-1.8cm}
\includegraphics[width=90mm]{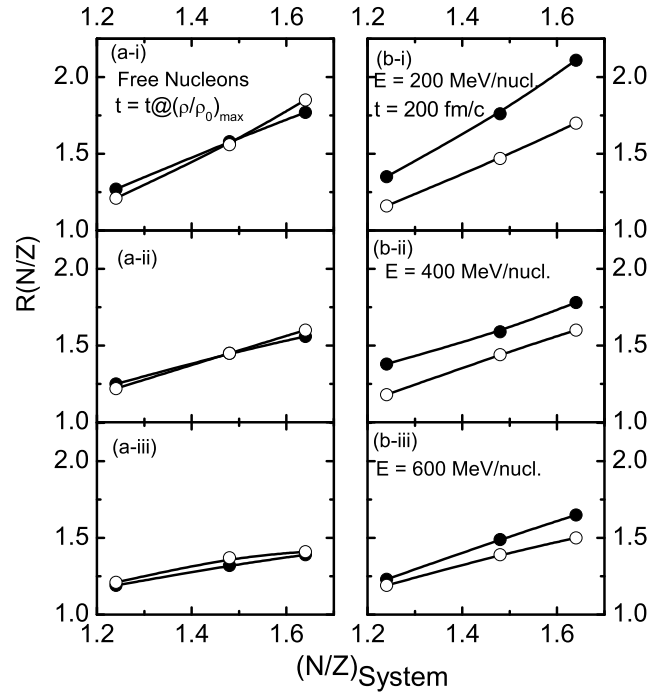}
\caption{\label{fig:5} Isospin asymmetry dependence
of the single neutron-to-proton ratio for free nucleons at different
incident energies. The left panel is at the time of maximum
compression, while the right panel is at the freeze-out time. Solid
and open circles represent the soft and stiff symmetry energies,
respectively. }
\end{figure}

\begin{figure}
\vspace{-2.0cm}
\includegraphics[width=90mm]{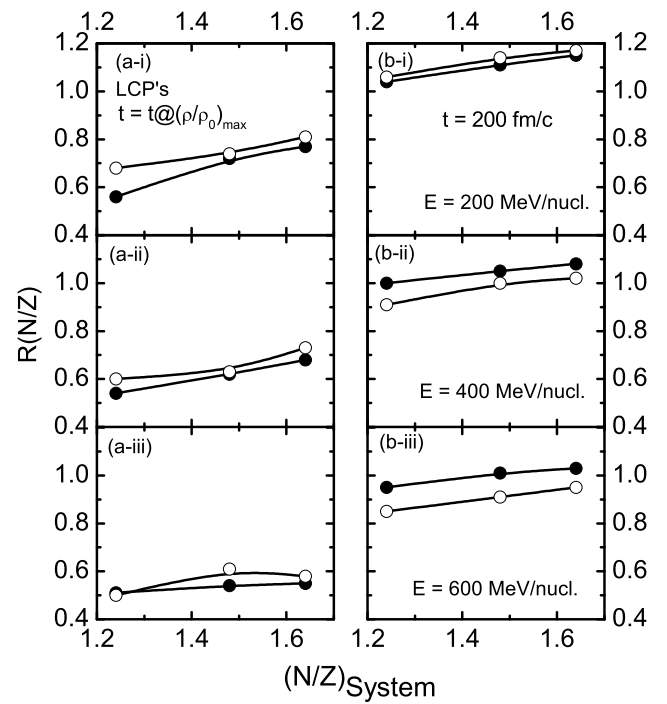}
\caption{\label{fig:6} Same as in Fig. \ref{fig:5}
but for the LCPs.}
\end{figure}

In order to address the sensitivity of symmetry energy at the time
of maximum compression and at freeze-out time,  we display the
incident energy and isospin asymmetry dependencies of the single
neutron-to-proton ratio at different times in Figs. \ref{fig:4}, \ref{fig:5} and
\ref{fig:6}.  The left and right panels are at the time
of maximum compression and freeze-out time, respectively.
In Fig. \ref{fig:4}, the ratios from free nucleons  and
LCPs  are displayed in the top and bottom panels. The many interesting
facts are revealed in the figure. The incident energy dependencies of
the ratios are found to be highly sensitive to symmetry energy for the
two different times. As we know, the relative strength of symmetry energy
is opposite at sub- and supra-saturation densities
with $\gamma_i=0.5$ and $\gamma_i=1.5$. In the range of 50-150
MeV/nucleon, only the low density part up to about 1.1$\rho_0$
contributes. Therefore, in the low-energy region, for free
nucleons, we can see the high ratio with the soft symmetry energy
at both the times  under consideration.
At and above 200 MeV/nucleon, a broad range of densities up to
1.8$\rho_0$ is involved. Of course, at  about 200 MeV/nucleon and
above, for the behavior of the high-density symmetry energy, there is
a combined effect for particles going through both low-and high-density region.
At higher energies, a higher $N/Z$ ratio is observed
with the stiff symmetry energy for the neutron rich system $^{132}$Sn
+ $^{132}$Sn at the time of maximum compression, which is
true with the soft symmetry energy at the freeze-out time. The result is
similar  for
free nucleons and LCPs. However, LCPs are not as sensitive and
the ratio is even less than the ratio of the system. This is
due to the excess number of protons involved in the production of
LCPs compared to free nucleons. These protons will lower
the ratio for LCPs.

It is clearly visible that the ratio at
both times is almost the same with the stiff symmetry energy, but
changes drastically with the soft symmetry energy. This is due to the
fact that, at the time of maximum compression, the density is in
the supra-saturation region and the stiff symmetry energy is much
higher (see Fig.\ref{fig:3}) than the soft symmetry energy. Therefore,
the stiff symmetry energy is able to separate most of the
neutrons near the time of maximum compression and then accelerate
the neutrons toward higher kinetic energy at later times.
However, the soft symmetry energy is not so high, and the
separation of the neutrons takes place for a longer time. After
50-60 fm/c (see Fig. \ref{fig:1}), the density drops to the
sub-saturation density region and now the soft symmetry energy has
a quite high magnitude (see Fig. \ref{fig:3}) compared to the
stiff one. The soft symmetry energy in this region is still
separating the neutrons as well as accelerating them high
kinetic energy. That is why the ratio with the soft symmetry
energy drastically changes when one goes from compression to
freeze-out time, but remains almost constant with the stiff
symmetry energy.

\begin{figure}
\vspace{-1.0cm}
\includegraphics[width=90mm]{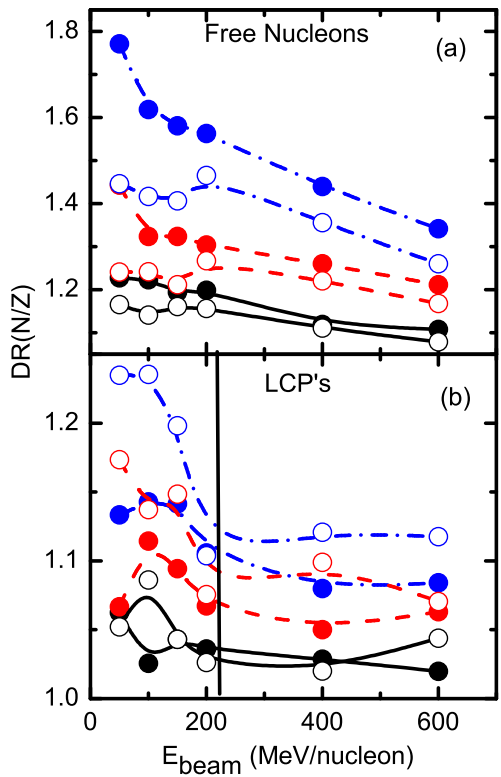}
\caption{\label{fig:7}(Color online) Excitation function
 of the double neutron-to-proton ratio from different
isotopes of Sn for free nucleons (top) and LCPs (bottom).
The vertical line in the bottom panel represent the energy limit above
which $DR(N/Z)$ of LCPs becomes more or less insensitive.
Solid and open circles represent the soft and stiff symmetry
energies, respectively. The solid, dashed, and dot-dashed line
corresponds to double ratios from $^{132}$Sn + $^{132}$Sn to
$^{124}$Sn + $^{124}$Sn, $^{124}$Sn + $^{124}$Sn to $^{112}$Sn +
$^{112}$Sn, and  $^{132}$Sn + $^{132}$Sn to $^{112}$Sn +
$^{112}$Sn, respectively. }
\end{figure}

Mainly, the neutron-to-proton ratio is found to decrease with the
incident energy for free nucleons as well as for LCPs, just like the
$\pi^-/\pi^+$ ratio.  The decrease in the ratio may be due to two
reasons:
\begin{itemize}
\item{One reason may be the role of Coulomb interactions with
incident energy. With increasing incident energy, chances of break-up of
initial correlations among the nucleons becomes stronger, and the production
of free nucleons including neutrons and protons will increase . However, at very low incident energy, the
production of neutrons is more due to the symmetry energy because of its repulsive (attractive)
nature for neutrons (protons).  In short, due to Coulomb interactions, a shift of
protons takes place from low to high incident energies. The effect of the Coulomb
interactions can be checked by taking the double ratio, which is
discussed in Fig. \ref{fig:7}}.
\item{The  contribution of pions from secondary-chance
nucleon-nucleon collisions might increase with the beam energy. If
a first-chance nucleon-nucleon collision converts a neutron to a
proton by producing a $\pi^-$, then subsequent collisions of the
energetic protons can convert them back to neutrons by producing
a  $\pi^+$. Therefore, at sufficiently high energy, the neutrons,
which are produced due to symmetry energy, are changing into the
protons and further producing $\pi$'s, which will lead to a decease in the
neutron-to-proton ratio. This can be confirmed by using the double
ratio concept. If the double ratio is still deceasing with
incident energy, then it means that, in addition to the Coulomb
interactions, the phenomenon of secondary nucleon-nucleon
collisions is also very important.}
\end{itemize}

One more point of interest is that  the difference between the
soft and stiff symmetry energies at freeze-out time is found to
decrease with incident energy for free nucleons, while it
increases  for LCPs. Of the above two reasons, the first one is
applicable for free nucleons as well as for LCPs. The second one
is applicable only for free nucleons, as the energy in this study
is up to 600 MeV/nucleon, which is quite sufficient to produce
pions.

To see the effect of the high-density behavior of symmetry energy on
the isospin asymmetry dependence, we display the ratio from free
nucleons and LCPs in Figs. \ref{fig:5} and \ref{fig:6} at only
high energies (200, 400, and 600 MeV/nucleon). Due to the instability of
the highly compressed zone, we are not able to differentiate
between the results of symmetry energy obtained at the time of
maximum compression; however, we had earlier obtained some
important conclusions from Fig. \ref{fig:4}, where incident
energy dependence was discussed.

The results from Figs. \ref{fig:5} and \ref{fig:6} at the freeze-out
time reveal many important points. The
isospin asymmetry dependence of the ratio from free nucleons is
highly sensitive to the symmetry energy compared to LCPs, i.e.,
the ratio from free nucleons is found to be sharply increasing
with the isospin asymmetry of the system compared to LCPs. This is
due to the fact that isospin effects on the ratio from free
nucleons are strongly affected by the symmetry energy and weakly
affected by Coulomb interactions, while the opposite is true for the
ratio from LCPs. As discussed earlier, the ratio is found to
decease with the incident energy, which is also true here for the isospin
asymmetry dependence.

The difference between the soft and
stiff symmetry energy results comes from the behavior of free nucleon
emission with the isospin asymmetry of the system, i.e., the greater
the isospin asymmetry of the system, the greater the contribution
of neutrons in the ratio due to the symmetry energy. The soft
symmetry energy is stronger at the freeze-out time, which will lead
to an increase in the ratio more sharply than the stiff
symmetry energy. This effect is again weakly observed in the ratio
from LCP's.
\begin{figure}
\vspace{-2.0cm}
\includegraphics[width=90mm]{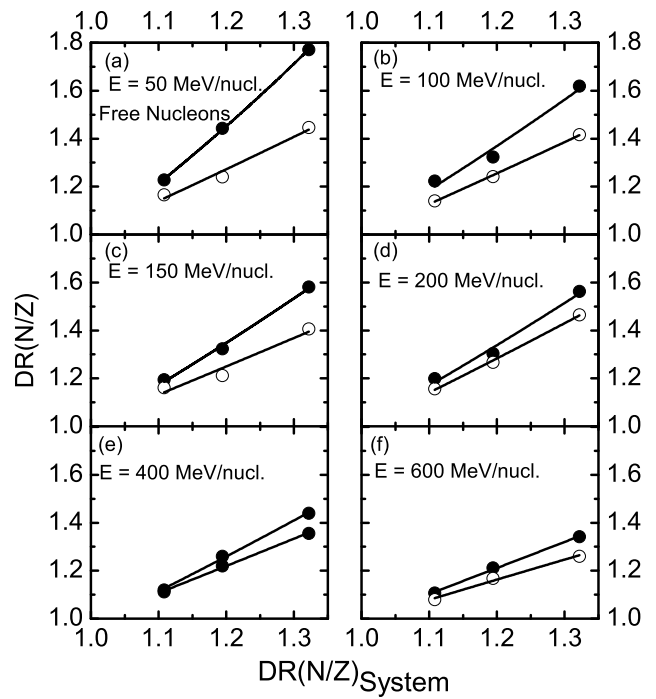}
\caption{\label{fig:8} Isospin asymmetry dependence
of the double neutron-to-proton ratio from free nucleons at different
incident energies. The different symbols have the same
meaning as in Fig. \ref{fig:5}. }
\end{figure}
\begin{figure}
\vspace{-1.9cm}
\includegraphics[width=90mm]{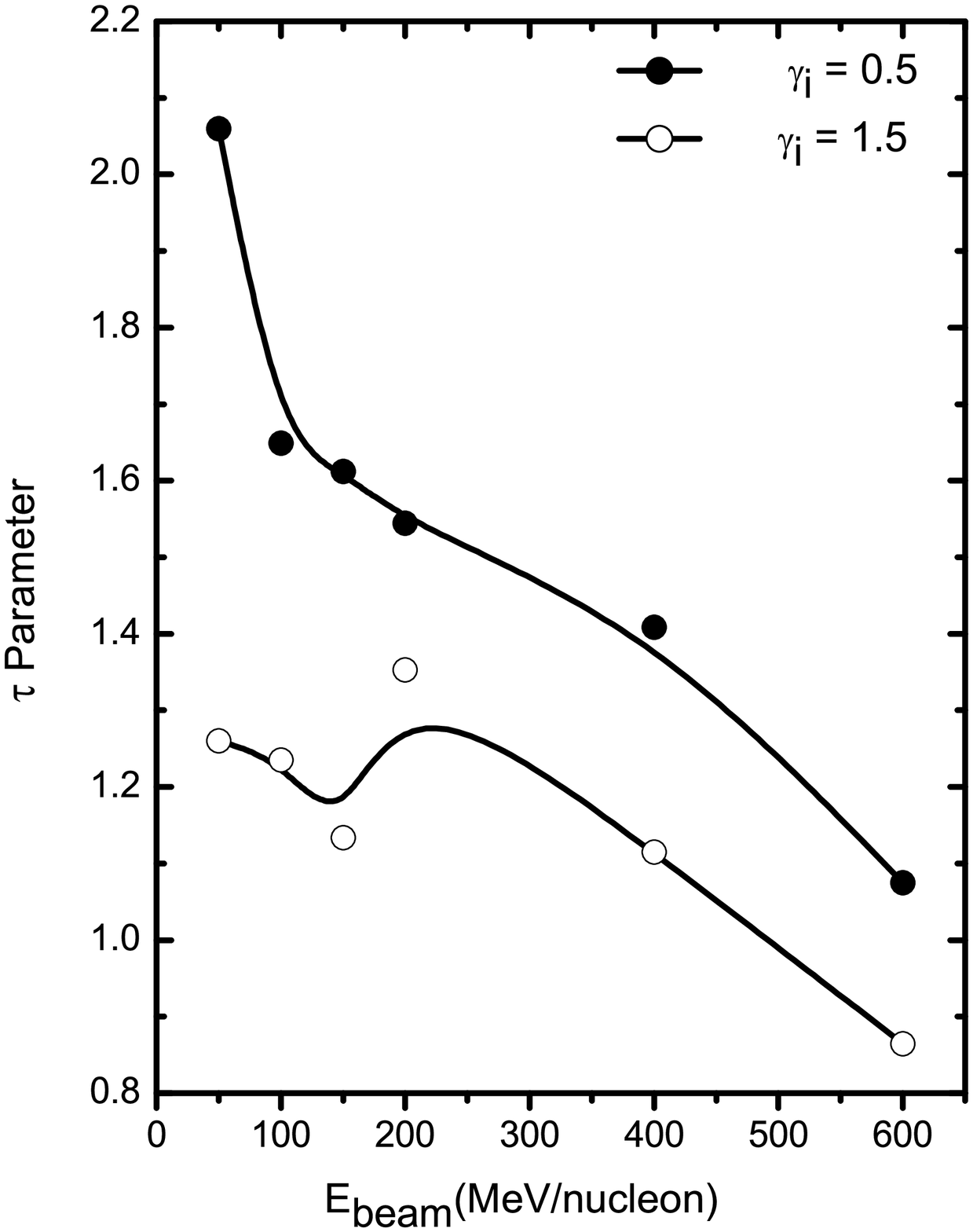}
\vspace{-1.0cm}\caption{\label{fig:9} Incident energy dependence of
the power-law exponent $\tau$ from Fig. \ref{fig:8}. The symbols
and lines are the same as in Figs.~5 and 8.}
\end{figure}

\subsection{Incident energy and isospin asymmetry dependencies of double neutron to proton ratio}
In order to cancel the Coulomb effects and to see the effect of
symmetry energy, we show in Fig. \ref{fig:7}, the incident
energy dependence of the double ratio from different isotopes of
Sn with different combinations,
namely, $^{132}$Sn + $^{132}$Sn and $^{124}$Sn + $^{124}$Sn,
$^{124}$Sn + $^{124}$Sn and $^{112}$Sn + $^{112}$Sn, $^{132}$Sn +
$^{132}$Sn and $^{112}$Sn + $^{112}$Sn, having differences of
8, 12, and 20 neutrons and the same number of protons. The upper
and lower panels are for free nucleons and LCPs. The
double ratio is found to increase with the difference of the number of
neutrons in the different combinations. It is similar to the
results obtained at sub-saturation densities by different models
\cite{Li97,Zhan08}. The main point to be discussed here is that
the double ratio is found to decrease with increasing
incident energy. As we have discussed in Fig. \ref{fig:4},
there may be two reasons for the decrease of the single ratio
with increasing incident energy. One was the Coulomb
effect, which is canceled out here. The second was the pion
effect, which is still active in the double ratio and
becomes more and more dominant with increasing incident
energy. Due to that effect, the double ratio is found to decrease with the
incident energy. It indicates that the pion production effect is very
important at high incident energy and is equally useful for
understanding the high-density behavior of the incident energy
\cite{Li09,Feng10}.

In contrast, this effect is valid only for the double ratio from
the free nucleons and not from the LCPs. The double ratio from
LCPs is found to be constant above 200 MeV/nucleon. This indicates
that the effect of the symmetry energy for the ratio from LCPs can be
analyzed only near sub-saturation densities close to
1.1$\rho_0$. The decrease in the single ratio for the LCPs was
only due to the Coulomb interactions at higher incident energies,
which is canceled out by taking the double ratio; the double
ratio from the symmetry energy becomes independent of the incident
energy after 200 MeV/nucleon. This type of dependence for the
single $\pi^-/\pi^+$ ratio can be observed above 1 GeV/nucleon
\cite{Li03}. The behavior of symmetry energy for the double ratio
is exactly the same as that for the single ratio. This indicates
that LCP production is also not a sensitive probe for
investigating the high-density behavior of the symmetry energy.
The only possible probe from the fragments is the double ratio of
neutrons to protons from free nucleons. Another possible probe is
the $\pi^-/\pi^+$ ratio, which recently was compared with the
experimental data of the FOPI by the IBUU04 and ImIQMD
calculations \cite{Li09,Feng10}.

In order to strengthen our conclusion, in Fig. \ref{fig:8}, we
display the isospin asymmetry dependence of the double ratio from free
nucleons at different incident energies.
 All the curves are fitted with a power law of the
form $y ~=~ax^{\tau}$, where $y$ is the double ratio from free nucleons
and $x$ is the double ratio of the systems. The power-law exponent
$\tau$ is found to vary drastically with the symmetry energy,
which is to be discussed later in Fig.
\ref{fig:9}. After canceling the Coulomb effects, the trend for
the double ratio is the same as that of the single ratio in Fig.
\ref{fig:5}. It reflects the fact that the isospin effects for free
nucleons is stronger for more neutron-rich systems and is mainly due
to the symmetry energy. However, the decrease in the isospin
effect with the increase of incident energy is due to the
production of pions at sufficiently high energy.
The difference in the double ratio obtained with the soft and
stiff symmetry energies here is also found to increase from the neutron-poor
to the neutron-rich system, just like the single neutron-to-proton
ratio in Fig. \ref{fig:5} as well as the single pion ratio in the
literature \cite{Li09,Feng10}.

\subsection{Incident energy dependence of Power law exponent $\tau$}
To see the clear systematics of the incident
energy toward the symmetry energy, we plot the incident energy
dependence of the power exponent $\tau$ in Fig.~\ref{fig:9}, which
is extracted from the curves of Fig.~\ref{fig:8}. With increasing
incident energy, the sensitivity of the symmetry energy goes on
decreasing toward the double ratio; however, the soft symmetry
energy is more sensitive in comparison with the stiff one.
In brief, when one goes from
the sub-saturation to the supra-saturation density region, the soft
symmetry still has a crucial role to play compared to the stiff one.
This is due to the density (Fig. \ref{fig:1}), which undergoes
a sudden change between the supra- and sub-saturation density regions
with time at higher incident energies.

Finally, from this study, we confirm that the high-density
behavior of symmetry energy can be studied by using the single and
double ratios of neutrons to protons from free nucleons. In
comparison, the double ratio is more accurate for this purpose,
due to its greater sensitivity to the  soft symmetry energy.
Meanwhile, the lighter and heavier fragments ratio can be
considered good candidates at sub-saturation densities, and
also have been  used in the literature many times by different
groups \cite{Li97,Zhan08}.

\section{Conclusion}
In order to investigate the high-density behavior of the symmetry
energy, isospin asymmetry and beam energy dependencies of neutron-to-proton
ratios (single and double) from different kinds of
fragments are studied by using the IQMD model. The single neutron-to-proton
ratio from free nucleons and LCPs is found to decrease
(increase)
with incident energy (with the isospin asymmetry of
the system). Stronger isospin effects are observed with the soft
symmetry energy. Similar results with the $\pi^-/\pi^+$ ratio are also
observed by Li {\it et al.} and Feng {\it et al.}, but with
opposite behavior for symmetry energy. The double neutron-to-proton
ratio from free nucleons is highly sensitive to the
symmetry energy, incident energy, and isospin asymmetry of
the system. However, the sensitivity of the neutron-to-proton double
ratio from LCPs to the nuclear symmetry energy is almost beam-energy
independent above 200 MeV/nucleon. The same trend is observed for
the single $\pi^-/\pi^+$ ratio above 1 GeV/nucleon. The sensitivity of
the soft symmetry energy to the ratio parameter is strongly
affected by the choice of times, which is not
true for the stiff symmetry energy. In simple words, just like
the $\pi^-/\pi^+$ ratio, the neutron-to-proton double ratio from free
nucleons can act as a useful probe to constrain the high-density
behavior of symmetry energy. Experiments are planned at MSU,
GSI, RIKEN, and FRIB to determine the high-density behavior of
symmetry energy by using the neutron-to-proton ratio.

\begin{acknowledgments}
This work is supported in part by the Chinese Academy of Sciences
Support Program for young international scientists under Grant
No. 2010Y2JB02, the  National Science Foundation  of China under Contract
No. 11035009, and No. 10979074,  by the the Knowledge Innovation Project of the
Chinese Academy of Sciences under Grant No. KJCX2-EW-N01, and by the 973-Program
under Contract No. 2007CB815004.
\end{acknowledgments}

\end{document}